%% file: main.tex
\documentclass[twoside,11pt]{article}

\usepackage{jmlr2e}
\usepackage[acronym,shortcuts,smallcaps,nowarn,nohypertypes={acronym,notation}]{glossaries}
\usepackage{amsmath}
\usepackage{amsfonts}       
\usepackage{amsthm}       
\usepackage{amsfonts}       
\usepackage{multirow}
\usepackage{graphicx}
\usepackage{subfigure}
\usepackage{bbm}
\usepackage{siunitx}
\usepackage[figuresright]{rotating}


\input{preamble/preamble_acronyms}

\ShortHeadings{AlphaMLDigger: A Novel Machine Learning Solution to Explore Excess Return on Investment}{Jimei Shen, Zhehu Yuan, Yifan Jin}
\firstpageno{1}

\begin{document}

\title{AlphaMLDigger: A Novel Machine Learning Solution to Explore Excess Return on Investment}


\author{\name Jimei Shen \email jimei.shen@cs.nyu.edu \\
      \addr Department of Computer Science\\
      New York University\\
      New York, NY 10012
      \AND
      \name Zhehu Yuan \email zy2262@nyu.edu \\
      \addr Department of Computer Science\\
      New York University\\
      New York, NY 10012
      \AND
      \name Yifan Jin \email yj2063@nyu.edu \\
      \addr Department of Computer Science\\
      New York University\\
      New York, NY 10012}

\editor{Rajesh Ranganath and Mark Goldstein}

\maketitle

\begin{abstract}
How to quickly and automatically mine effective information and serve investment decisions has attracted more and more attention from academia and industry. And new challenges have arisen with the global pandemic. This paper proposes a two-phase AlphaMLDigger that effectively finds excessive returns in a highly fluctuated market. In phase 1, a deep sequential natural language processing (NLP) model is proposed to transfer Sina Microblog blogs to market sentiment. In phase 2, the predicted market sentiment is combined with social network indicator features and stock market history features to predict the stock movements with different Machine Learning models and optimizers. The results show that the ensemble models achieve an accuracy of 0.984 and significantly outperform the baseline model. In addition, we find that COVID-19 brings data shift to China's stock market. 
\end{abstract}

\section{Introduction} \label{intro}

By fundamental financial hypnosis, in a perfectly efficient market, all information will be quickly and correctly understood by the market participants and will immediately be reflected in the market prices \cite{fama1970efficient}. However, the price of the stock does not always equal its intrinsic value. There are arbitrage opportunities because people are bounded rational \cite{simon1955behavioral} and there is an information cost \cite{grossman1980impossibility}. How to effectively extract useful information and make wise investment decisions remain a hard nut. Several influential attempts at asset pricing by history information have been widely used, such as Capital Asset Pricing Model \cite{sharpe1964capital}, Black–Scholes Model \cite{black1973pricing}. But these traditional methods have challenges in solving high-dimensional stochastic control problems in big data scenarios. Machine learning (ML) is a natural fit for digging excess returns (alpha) in the stock market because of its ability to automatically process massive amounts of structured or unstructured data, learn from history, and make predictions. Then comes the question: Which ML method(s) should we use to solve the stock prediction task?

Sentiment analysis and its main applications, known as the task of exploring authors' opinion of a given entities, is one of the hottest research area in information retrieval \cite{feldman2013techniques}. In sentiment analysis, the opinion to a specific piece of text is typically divided into three classes - positive, negative, and neutral. In the meanwhile, here are three scenarios of stock price movements - up, down, and stay the same. However, stock price rarely remains exactly the same. Therefore, it can be further simplified to an up-to-down binary sentiment classification task. Early pioneers have tried to apply sentiment analysis in stock trend prediction. Building such a model is extremely challenging, but its precision and applicability are very promising \cite{wuthrich1998daily}. Mittermayer (2004) proposes NewsCATS, an intraday stock forecasting model with text mining techniques \cite{mittermayer2004forecasting}. NewsCATS demonstrated that ML techniques can learn and choose the right way to trade, and there is much room for improvement. Early studies show that the Support Vector Machine (SVM) performs better than other ML techniques in this task \cite{schumaker2009quantitative, groth2011intraday}. A more recent study finds that the tree-based Decision Tree model outperforms SVM \cite{bogle2015sentamal}. With the evolution of tree-based models, more efficient models, such as Random Forest \cite{ho1995random}, XGBoost (XGB) \cite{chen2016xgboost}, LightGBM (LGB) \cite{ke2017lightgbm}, have been proposed and widely applied in other areas such as computer vision, pattern recognition, and natural language processing. The power of such tree-ensemble models in the finance task is underexplored. The models mentioned above solve the question of how to use sentiment indicators to predict stock trends. But how can we convert stock-related text into sentiment representatives?

Early financial text information retrieval methods are highly statically based, such as the product of term frequency (TF) and inverse document frequency (IDF) \cite{mittermayer2004forecasting}, bags of words, noun phrases and named entities \cite{schumaker2009quantitative}, financial sentiment dictionary \cite{cabanski2017hhu}. In recent years, the use of deep learning to automatically learn features has gradually replaced artificially constructed features and statistical methods. But one of the key issues is the huge amount of data required. Inspired by human behavior, instead of learning from scratch, we can use prior knowledge and a relatively small number of special cases to solve problems. Transfer learning thus becomes mainstream in natural language processing \cite{devlin2018bert}. Transfer learning consists of two parts: pre-training and fine-tuning. Pre-training obtains task-independent models from large-scale data through self-supervised learning, that is, the semantic representation of a word in a specific context. The second step is fine-tuning, which corrects the pre-trained model for a specific task. Araci (2019) first applies pre-training to the financial sentiment analysis task and proposed FinBERT. FinBERT finds that the pre-training method is superior to other methods \cite{araci2019finbert}.

However, the methods mentioned above are mainly aimed at the US stock market. Whether these methods are effective for other markets, such as the second largest economy in the world? Furthermore, the unprecedented COVID-19 dramatically changed the stock market \cite{mazur2021covid}. Does the performance of these methods affected by COVID-19? How to extract useful financial information and take action one step ahead for alpha for now? To answer these questions, this paper proposes AlphaMLDigger. The main contributions of this paper are: (1) We found a significant reduction on model predictability after COVID-19 outbreak, which indicates an decrease on China's market efficiency. (2) Our proposed deep sequential modeling method achieved outstanding results on Chinese financial text sentiment analysis task. (3) Our method is robust to the negative influence of COVID-19 on market efficiency to some extent.

\section{Related Work} \label{related}

\hspace{1.5em}\cite{de2008can} used the text information for the Engadget communication to predict stock prices. The experimental results show that the Support Vector Machine (SVM) could effectively predict changes related to stock prices in blog posts (with a precision of 78\% to 87\%. This work specifically looked at the impact on stocks of technology companies due postings in a gadget-discussing blog without including any Chinese companies. 

\cite{schumaker2009quantitative} built a SVM-based Arizona Financial Text System (AZFinText) to solve the problem of discrete forecasting of stock prices with financial news. The system has 54.6\% accuracy and is traded with a return of 8.50\%, 2\% higher than the best performing quant fund at that time. Furthermore, \cite{tan2019tensor} proposed the tensor-based eLSTM model in Chinese financial news to predict stock movements. The experiments showed that the eLSTM (67.1\% accuracy) outperforms existing algorithms such as AZfinText, eMAQT, TeSIA, and LSTM. However, the accuracy of these works is significantly lower than our result.


\cite{zhang2012method} conducted text sentiment analysis on Sina Microblog related to stock topics and proposed a method to automatically identify investors' sentiment. Compared with SVM, NB, Maximum Entropy (ME) sentiment polarity classifier has the highest accuracy (85.7\%). This work collects data from both normal users and professional users, while our work focuses on six officially verified bloggers who are usually considered to be more professional. In addition, \cite{zhang2012method} lacks data after the COVID-19 situation, which has a significant effect on the stock market, as shown in our paper.


\cite{bogle2015sentamal} proposed SentAMaL, which studied the impact of the sentiment expressed on Twitter on US stock indexes. The accuracy of the Decision Tree (DT) reached 100\% on the Combined Index prediction. Tree models outperformed SVMs and NNs in this task. This work especially looks at the US stock market, which is significantly different from the Chinese stock market.




\section{Methodology}
\subsection{Data}
We use a web crawler to collect stock related blogs from Sina Microblog and A-share index (000002.SH) price and volume data. The blogs are from 6 official vertified Sina Microblog financial bloggers. Price and volume data are from Sina Finance.

Our aim is to predict the movements of stock price. We define the next 1 day return of the price $r_t = \frac{p_{t+1}-p_t}{p_t}$, where $p_t$ is the closing price of the A-share index at time $t$. If $r_t > 0$, we define the target as $1$, otherwise $0$. We divide the data into three datasets. In this way, we can find out the impact of COVID-19 to China's stock market and the robustness of our method.

\begin{itemize}
    \item \textbf{Before COVID-19 1}. 12,427 micro-blogs posted from 2018-01-01 to 2018-12-31 are collected. This dataset is for word-to-sentiment task. \textit{Train}: 90\% of the data are used as a training set for a natural language processing (NLP) model. \textit{Test}: the rest 10\% of the data are used to test the performance of the NLP model.
    
    \item \textbf{Before COVID-19 2}. 15,266 microblogs posted from 2019-01-01 to 2019-12-31 are collected. This dataset is for the stock movements prediction task. The blog contents are fed into the NLP model trained before to get the sentiment as an input feature. \textit{Train}: 80\% of the data are used as a training set for the stock prediction models. \textit{Test1}: the rest 20\% of the data are used to test the performance of the stock prediction models before COVID-19.
    
    \item \textbf{During COVID-19}. 2,302 microblogs posted from 2020-01-01 to 2020-03-06 are collected. This period was when COVID-19 began in China. The blog contents are fed into the NLP model trained before to get the sentiment as an input feature. \textit{Test2}: the data is 100\% used to test the performance of the stock prediction models during COVID-19.
\end{itemize}

\subsection{Model}

\cite{cho2014learning} proposed the Gated Recurrent Unit (GRU). GRU is a very effective variant of LSTM often used by predecessors in this task. In addition to solving the RNN gradient disappearance and explosion problem, the more straightforward structure of GRU makes it compute faster. For time $t$, weight metrics $w$, bias parameters $b$, input $x_t$, hidden state $h$, reset gate $r$, update gate $z$, Sigmoid function $\sigma$, the GRU cell calculate as Equation \ref{e1} to \ref{e4}.
\begin{equation}
    r_t = \sigma(w_r \cdot [h_{t-1}, x_t]+b_r) \label{e1}
\end{equation}
\begin{equation}
    z_t = \sigma(w_z \cdot [h_{t-1}, x_t]+b_z)
\end{equation}
\begin{equation}
    \widetilde{h}_t = tanh(w_h \cdot [r_t\ast h_{t-1},x_t]+b_h)
\end{equation}
\begin{equation}
    h_t=(1-z_t)\ast h_{t-1}+z_t \ast \widetilde{h}_t \label{e4}
\end{equation}

\begin{figure}[!htbp]
\centerline{\includegraphics[scale=0.25]{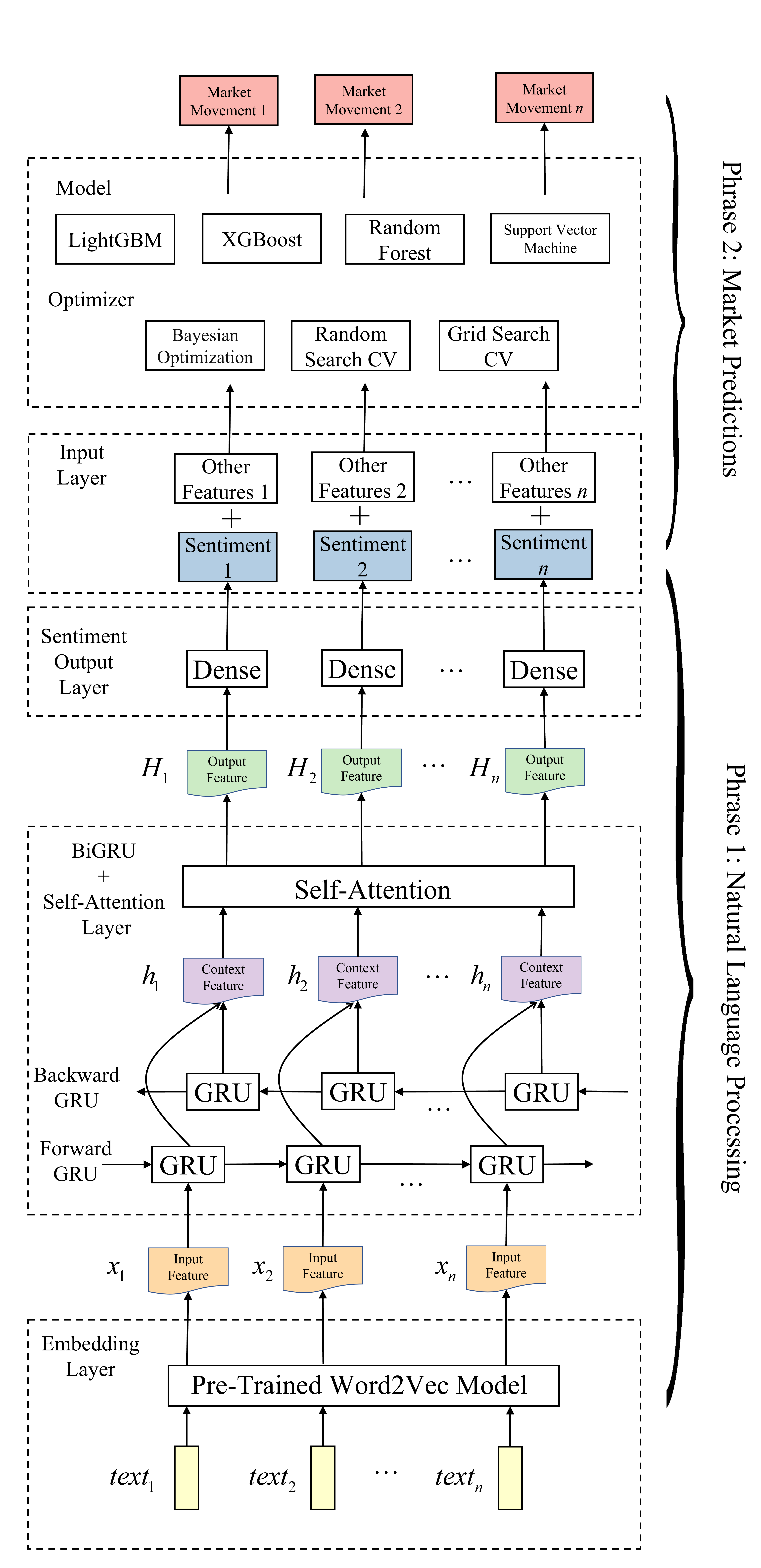}}
\caption{AlphaMLDigger Model Architecture.}
\label{fig}
\end{figure}

To capture the context features of financial news, we use BiGRU. At time $t$, the hidden layer of the forward GRU unit has $\overrightarrow{h_t} =GRU(x_t, \overrightarrow{h_{t-1}} )$ and the backword GRU unit $\overleftarrow{h_t} =GRU(x_t, \overleftarrow{h_{t-1}} )$, then $h_t = [\overrightarrow{h_t}, \overleftarrow{h_t}]$. To capture important information, self-attention is used. $H_t = m(h_t)$, where $m(h_t) = concat \left ( attention(h_t), attention(h_t), \cdots, attention(h_t) \right)W^{O}$, $attention$ is the Multiplicative Attention. Finally, a fully-connected is added for output.

SVM is commonly used in this task, as mentioned in Sections \ref{intro} and \ref{related}. Therefore, we consider SVM as our baseline model. Unlike SVM \cite{noble2006support}, which is a largest margin linear classifier in the feature space, RF \cite{ho1995random}, XGB \cite{chen2016xgboost}, LGB \cite{ke2017lightgbm} are ensemble models of DT using different parallel schemes, loss of gradient computation, feature processing methods, etc. Random Search, Grid Search, Bayesian Optimization (BO) are commonly used methods for hyperparameter optimization in machine learning. Random and Grid Search are to select the optimal parameter combination from parameter spaces. Grid Search traverses all parameter space while Random Search just samples several parameter combinations. \cite{bergstra2012random}. Both Grid and Random Search ignore the information of the previous point when testing a new point. In contrast, BO finds a balance between exploration of unknown points and exploitation of where the global maximum value is most likely to occur according to the posterior distribution \cite{frazier2018tutorial}. Cross-validation is a practical way to reduce overfitting problems. First, the training set is divided into $k$ parts, and the parameter space of model is defined. For each model composed of randomly selected parameter combinations, the $j$-th part is selected in the $k$ parts of the data set for verification, and the rest are used for training. Test the training parameters on the $j$ th data to get the test error. The model with the smallest sum of all errors is our optimal model. Our AlphaMLDigger model is designed as shown in Figure \ref{fig}.

\section{Experiments}

\begin{figure}[!htbp]
\centerline{\includegraphics[scale=0.3]{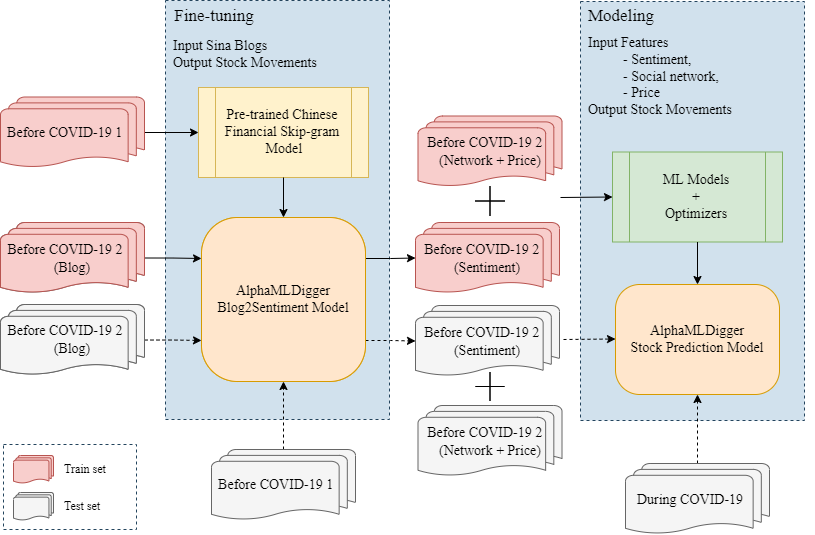}}
\caption{AlphaMLDigger System Design.}
\label{alphamldigger}
\end{figure}

Our experiment platform is NYU HPC. We use a Linux system with 14 cores CPU and 64 GB memory, Miniconda 3 with Python 3.7. Our AlphaMLDigger system is designed as shown in Figure \ref{alphamldigger}.

The first step is to convert the post contents in Chinese into sentiment, so that we could build machine learning models with it as an input feature. First, we remove the punctuation marks of each sample and then use Jieba word segmentation. Then we convert the word segmentation result into a list and index it, so that each news becomes an index number corresponding to the word in the pre-trained word vector. This project uses pre-trained word + ngram financial news context features trained with word2vec/skip-gram with negative sampling \cite{P18-2023}. To save computation power, speed up training process, and prevent overfitting, we take a compromise length $max\_tokens = mean(num\_tokens) + 2 \ast std(num\_tokens)$, where $num\_tokens$ is the list contains all training tokens. We choose $num\_tokens=134$ which can cover  97.29\% data sample. Finally, we do padding and truncating. We find the very beginning of the blogs are usually less meaningful tags. Thus, we use the 'pre' method, which pads the front with 0 on which length is shorter than 134 and truncates the front on which length is longer than 134. We set the 7 layers of BiGRU with 256, 128, 64, 32, 16, 8, 4 units followed by self-attention and dense with Sigmoid activation. We feed the train pad into BiGRU with Attention on the top on the 90\% training set, the remaining 10\% testing set shows that the sentiments have 82.14\% accuracy.

\setlength\tabcolsep{2.5pt}
\begin{table}
\centering
\caption{Data Sample}\label{data}
\begin{tabular}{llllllll} 
\hline
sentiment & T & C & F & high     & low      & open     & label  \\ 
\hline
0.305082  & 258      & 78        & 112        & 3408.315 & 3327.868 & 3386.851 & 0      \\
0.312336  & 27      & 5        & 6        & 3430.720 & 3393.436 & 3404.296 & 1      \\
0.877553  & 31      & 14        & 10        & 3066.186 & 2989.352 & 2989.352 & 1      \\
\hline
\multicolumn{8}{l}{$^{\mathrm{a}}$where T, C, F stands for thumbs ups, comments, and forwards.}
\end{tabular}
\end{table}

The second step is to predict the movements of the stocks. First, feed two new datasets (before and during COVID-19) into the NLP model trained in the first step and get the sentiment feature from blogs. As shown in Table \ref{data}, the input characteristics are (1) the sentiment of the blog contents; (2) thumbs, comments, and forwards show the influence of the blog; (3) high, low, and open are the history data from the previous day. The label is our target that indicates stock movements (0 for not raise and 1 for raise). The result comparing 4 models combined with 3 optimizations is shown in Table \ref{result}. The accuracy increase to 98\% with LGB and XGB, and 97\% with RF. However, using SVM, we get a worse accuracy. Because SVM uses a single linear kernel that cannot learn the pattern behind sequential data. For the first 3 models, the optimization method does not make a significant difference. But in SVM, BO performs much better than Grid and Random Search. We could observe that after COVID-19 outbroke, the accuracy is significantly reduced for every method.
 
\begin{table}
\centering
\caption{Test Results}\label{result}
\begin{tabular}{ll|lll} 
\hline
Model                & Items & BO & Random
 & Gird                           \\ 
\hline
\multirow{3}{*}{LGB} & Running time (s)       & 140.573&	21.116	& 412.649
                                                               \\
                     & Test1 Accuracy    & 0.983&	0.984 &	0.981                          \\

                     & Test2 Accuracy  & 0.729	&0.748&	0.728
   \\
\hline
\multirow{3}{*}{XGB} & Running time (s)       & 1429.758&	209.824	 &6779.928

                                                               \\
                     & Test1 Accuracy   & 0.981&	0.982&	0.985
                         \\

                     & Test2 Accuracy  & 0.729	&0.748	&0.735

   \\
\hline
\multirow{3}{*}{RF} & Running time (s)       & 444.818	&17.442&	24.115

                                                               \\
                     & Test1 Accuracy    & 0.976&	0.975	&0.975
                   \\

                     & Test2 Accuracy  & 0.745	&0.754&	0.754

   \\
\hline
\multirow{3}{*}{SVM} & Running time (s)       & 424.223&	1.119&	1.256

                                                               \\
                     & Test1 Accuracy    & 0.792&	0.492	&0.509
                        \\

                     & Test2 Accuracy  & 0.720&	0.406&	0.594

   \\
\hline
\end{tabular}
\end{table}

\section{Conclusion \& Discussion}
\subsection{Trade-off between Running Time and Accuracy}

While Grid Search traverses every combination in parameter space, which guarantees a global optimal in the training set, it is very time consuming, especially for models with a large number of parameters and relatively poor parallel computing design, such as XGB. Random Search may miss the optimal solution, however, sometimes the non-optimal parameter on the training set can have a better generalization ability to the test set, as we can find from Test1 and Test2 results on LGB. Adding posterior knowledge in parameter optimization is a good idea. However, it also requires a lot of resources and time. For those high-dimensional, nonconvex functions with unknown smoothness and noise, the BO algorithm is often difficult to fit and optimize, as we can find from LGB and XGB. The effect is unstable because of the randomness of initialization. Its effect is not stable. We suggest using Random Search CV when the dataset and the parameter space are considerably large. When the parameter space is small with a linear model, we suggest BO.

\subsection{Negative Impact on China's Stock Market Efficiency}
Our proposed model can achieve higher accuracy on the test set before COVID-19 than existing related works. However, a notable decrease in models' predictability occurs with COVID-19 despite Test1 and Test2 being very close on time. Although the market efficiency is reduced, it creates opportunities for arbitrage and for abnormal returns \cite{dias2020random}. Our goal is to predict the buy signal (1). We can find from Table \ref{covidresult}, the precision of 1 is affected by COVID-19 much less than 0 and remains at an acceptable rate, which means that our method is robust to the negative impact of COVID-19 to some extent and has the ability to continuously earn alpha.

\begin{table}[!htp]
\centering
\caption{Test results before and after COVID-19 Outbroke} \label{covidresult}
\centering
\begin{tabular}{lll|lll|lll} 
\hline
\multicolumn{3}{l|}{Items}               & \multicolumn{3}{l|}{Test before COVID-19} & \multicolumn{3}{l}{Test during COVID-19}  \\ 
\hline
Model                & Optimizer & Label & Precision & Recall & F1                                & Precision & Recall & F1                               \\ 
\hline
\multirow{6}{*}{LGB} & \multirow{2}{*}{BO}        & 0     & 0.985     & 0.979  & 0.982                             & 0.690     & 0.602  & 0.643                            \\
                     &           & 1     & 0.980     & 0.986  & 0.983                             & 0.750     & 0.816  & 0.782                            \\
                     & \multirow{2}{*}{Random}    & 0     & 0.985     & 0.982  & 0.984                             & 0.720     & 0.618  & 0.665                            \\
                     &           & 1     & 0.983     & 0.986  & 0.984                             & 0.762     & 0.836  & 0.798                            \\
                     & \multirow{2}{*}{Gird}      & 0     & 0.983     & 0.979  & 0.981                             & 0.696     & 0.585  & 0.636                            \\
                     &           & 1     & 0.980     & 0.983  & 0.981                             & 0.744     & 0.826  & 0.783                            \\ 
\hline
\multirow{6}{*}{XGB} & \multirow{2}{*}{BO}        & 0     & 0.980     & 0.981  & 0.980                             & 0.689     & 0.605  & 0.644                            \\
                     &           & 1     & 0.981     & 0.981  & 0.981                             & 0.751     & 0.814  & 0.781                            \\
                     & \multirow{2}{*}{Random}    & 0     & 0.978     & 0.985  & 0.982                             & 0.712     & 0.636  & 0.672                            \\
                     &           & 1     & 0.986     & 0.979  & 0.982                             & 0.768     & 0.825  & 0.796                            \\
                     & \multirow{2}{*}{Gird}      & 0     & 0.987     & 0.982  & 0.984                             & 0.698     & 0.614  & 0.653                            \\
                     &           & 1     & 0.983     & 0.987  & 0.985                             & 0.756     & 0.819  & 0.786                            \\ 
\hline
\multirow{6}{*}{RF}  & \multirow{2}{*}{BO}        & 0     & 0.973     & 0.978  & 0.975                             & 0.719     & 0.611  & 0.661                            \\
                     &           & 1     & 0.979     & 0.974  & 0.976                             & 0.759     & 0.837  & 0.796                            \\
                     & \multirow{2}{*}{Random}    & 0     & 0.971     & 0.979  & 0.975                             & 0.736     & 0.614  & 0.669                            \\
                     &           & 1     & 0.979     & 0.972  & 0.975                             & 0.763     & 0.849  & 0.804                            \\
                     & \multirow{2}{*}{Gird}      & 0     & 0.971     & 0.979  & 0.975                             & 0.736     & 0.614  & 0.669                            \\
                     &           & 1     & 0.979     & 0.972  & 0.975                             & 0.763     & 0.849  & 0.804                            \\ 
\hline
\multirow{6}{*}{SVM} & \multirow{2}{*}{BO}        & 0     & 0.835     & 0.719  & 0.772                             & 0.811     & 0.404  & 0.539                            \\
                     &           & 1     & 0.760     & 0.862  & 0.808                             & 0.697     & 0.936  & 0.799                            \\
                     & \multirow{2}{*}{Random}    & 0     & 1.000     & 0.137  & 0.240                             & 0.944     & 0.018  & 0.036                            \\
                     &           & 1     & 0.545     & 1.000  & 0.706                             & 0.599     & 0.999  & 0.075                            \\
                     & \multirow{2}{*}{Gird}      & 0     & 1.000     & 0.136  & 0.240                             & 0.944     & 0.018  & 0.036                            \\
                     &           & 1     & 0.545     & 1.000  & 0.706                             & 0.599     & 0.999  & 0.749                            \\
\hline
\end{tabular}
\end{table}

\subsection{Limitation \& Future Work}
First, the dataset is limited on 6 bloggers in Sina Microblog, while a more general dataset from more bloggers and social media may get a more reliable result. More resources should also be considered, such as financial news. Second, The number of supervised learning models and optimizations is limited. More methods are expected in the future. Third, for the choices of $y$, instead of daily return, we can also consider a higher frequency of trading, such as in minutes or seconds, or a longer return, such as 3-days or 7-days. Fourth, conditions on the sale side. We propose a long-only strategy and define when to buy and keep how long. We take the sale side into account and develop buy-and-sell strategies.

\newpage

\acks{The authors would like to thank Professor Rajesh Ranganath and Mark Goldstein for their insightful comments and suggestions which motivates us to broaden our research in various aspects.}

\bibliography{sample}

\end{document}

%% file: preamble/preamble_acronyms.tex
\newacronym{KL}{kl}{Kullback-Leibler}

\newacronym{SVI}{SVI}{Stochastic Variational Inference}